\documentclass[doublecol,figures]{epl2}

\usepackage{color,amsmath,amsfonts}
\usepackage{mathrsfs}

\newcommand{\T}{\Theta}

\title{First passage behaviour of fractional Brownian motion in
two-dim\-ensional wedge domains}
\shorttitle{First passage of FBM in 2D wedge domain}

\author{Jae-Hyung Jeon\inst{1}\thanks{E-mail: \email{jae-hyung.jeon@ph.tum.de}}
\and Aleksei V. Chechkin\inst{2,3}
\and Ralf Metzler\inst{1,4}\thanks{E-mail: \email{metz@ph.tum.de}}}
\shortauthor{J.-H. Jeon \etal}

\institute{
\inst{1} Department of Physics, Technical University of Munich,
James-Franck Stra{\ss}e, 85747 Garching, Germany\\
\inst{2} Akhiezer Institute for Theoretical Physics NSC KIPT,
Akademicheskaya Str.1, 61108 Kharkov, Ukraine\\
\inst{3} School of Chemistry, Tel Aviv University, Ramat
Aviv, Tel Aviv 69978, Israel\\
\inst{4} Department of Physics, Tampere University of Technology,
FI-33101 Tampere, Finland}


\abstract{We study the survival probability and the corresponding first
passage time density of fractional Brownian motion confined to a
two-dimensional open wedge domain with absorbing boundaries. By
analytical arguments and numerical simulation we show that in the long
time limit the first passage time density scales as $\wp_\T(t)\simeq t^{-1+
\pi(2H-2)/(2\Theta)}$ in terms of the Hurst exponent $H$ and the wedge angle
$\Theta$. We discuss this scaling behaviour in connection with the reaction
kinetics of FBM particles in a one-dimensional domain.}

\begin{document}

\maketitle

\section{Introduction}

The first passage defines the moment at which a random quantity crosses a
given threshold value for the first time. For instance, this could be the
time when a random walker first leaves a defined region of space. Other
examples include phenomena as diverse as diffusion-limited reactions
\cite{smoluchowski}, cyclisation of a polymer \cite{igor,likthman}, stock
market dynamics \cite{bouchaud}, or search problems \cite{search}. The
concept of first passage statistics is of quite ubiquitous importance
in the theory of stochastic processes \cite{redner}.

First passage characteristics have been investigated for various geometries.
Thus, on semi-infinite domains the probability density of the first passage
generally
displays a power-law tail, and the associated mean first passage time diverges
\cite{redner}. Recently, remarkable progress has been achieved, relating the
underlying geometry to the resulting first passage behaviour \cite{benichou}.

In the case of normal diffusion, characterised by the linear growth $\langle
\mathbf{r}(t)^2\rangle\simeq t$ of the ensemble-averaged mean squared
displacement, the associated first passage problems are obtained by solving
the associated diffusion equation with given boundary conditions or, for
simpler geometries, by the method of images \cite{redner}. For anomalous
diffusion with the non-linear growth \cite{report}
\begin{equation}
\label{msd}
\langle\mathbf{r}^2(t)\rangle\simeq t^{2H},\,\,\,2H\neq1,
\end{equation}
of the mean squared displacement, no general theory exists to obtain the
corresponding first passage time statistics. Only for certain classes of
anomalous diffusion models analytical approaches for the first passage
properties are known. Examples include the continuous time random walk
model \cite{scher} governed by a heavy-tailed waiting time distribution
with diverging characteristic waiting time \cite{mekla2000,Rang,condamin2,%
condamin} or by a power-law jump length distribution with
diverging variance (L{\'e}vy flights) \cite{chechkin}; as well as
the diffusion on fractal media \cite{condamin3}. Importantly the
first passage time behaviour differs between different stochastic
processes despite sharing the same form (\ref{msd}) of the mean
squared displacement with dynamic exponent $2H$.

Fractional Brownian motion (FBM), originally introduced by
Kolmogorov \cite{kolmogorov} and later re-discovered by Mandelbrot
and van Ness \cite{vanness}, is a generalised Gaussian process,
whose increments in one spatial dimension,
\begin{equation}
dx(t)=\xi^H(t)dt,
\end{equation}
are stationary and normally distributed but dependent. Here the quantity $\xi
^H(t)$ represents fractional Gaussian noise with zero mean ($\langle\xi^H(t)
\rangle=0$) and autocorrelation
\begin{eqnarray}
\nonumber
\langle\xi^H(t_1)\xi^H(t_2)\rangle&=&2K_HH(2H-1)|t_1-t_2|^{2H-2}\\
&&\hspace*{-0.8cm}+4K_HH|t_1-t_2|^{2H-1}\delta(t_1-t_2).
\label{autocorrelation}
\end{eqnarray}
$K_H$ of dimension $[K_H]=\mathrm{cm}^2/\mathrm{sec}^{2H}$ is the
anomalous diffusion coefficient. The fractional Gaussian noise is
negatively (antipersistence) or positively (persistence) correlated
for the cases of subdiffusion and superdiffusion, respectively.
FBM has been widely used to describe anomalous diffusion
phenomena including annual river discharge \cite{hurst}, stock
market dynamics \cite{econo}, the motion of tracer particles in
crowded environments \cite{weiss,lene,magdziarz,weber}, or
single-file diffusion \cite{lizana}. Despite its popularity and
deceivingly simple definition in terms of fractional Gaussian
noise the exact stochastic properties of FBM are not well
understood. The first passage properties are known only for the
one-dimensional semi-infinite domain. The associated long-time
asymptotic form of the first passage time density $\wp(t)\sim
t^{H-2}$ was conjectured by Ding and Yang from scaling arguments
\cite{ding}, and rigorously proved by Molchan
\cite{molchan}. Recently, it was shown that the Wilemskii-Fixmann
approximation, originally introduced for the description of
polymer cyclisation, produces the first passage time behaviour and
barrier crossings of particles driven by fractional Gaussian noise
\cite{oleksii}. The difficulty to analytically
access FBM's first passage properties is related to the fact that
FBM is a strongly non-Markovian process, which does not fall into the
class of semimartingales \cite{weron}.

\begin{figure}
\begin{center}
\includegraphics[width=6.4cm]{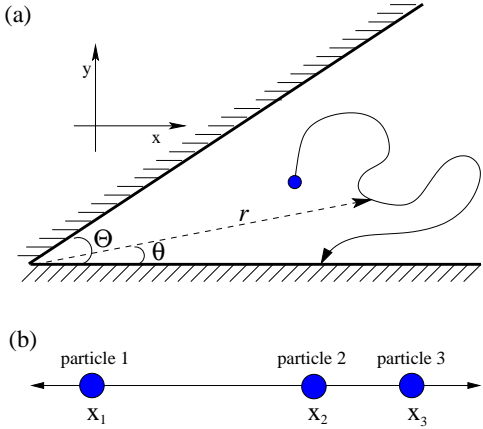}
\end{center}
\caption{(a) Stochastic motion of a particle confined to a two-dimensional
wedge of opening angle $\Theta$. The particle motion motion at time $t$ is
described by $\mathbf{r}(t)=(r(t),\theta(t))$ in polar coordinates. The
excursion is terminated when the particle hits one of the absorbing walls
located at $\theta=0$ or $\Theta$, the moment of first passage.
(b) Diffusion-limited reaction of three particles that move according to
FBM in one-dimensional space.}
\label{wedge}
\end{figure}

Here we analyse the first passage time behaviour of
two-dimensional FBM confined to a wedge domain of opening angle
$\Theta$ (Fig.~\ref{wedge}). This problem is of particular
interest as the reaction kinetics of three diffusive particles in
one-dimensional space can be mapped on this problem
\cite{redner,avraham}. In the case of ordinary Brownian motion
($H=1/2$), the wedge problem can be solved exactly using the
Green's function formalism \cite{redner}. Based on analytical and
numerical arguments we conjecture the first passage behaviour of
FBM for arbitrary Hurst exponent in a wedge geometry.

In the following Section we define multi-dimensional FBM and the
first passage in a wedge domain. We then review the Green's
function formalism for the Brownian case and obtain
analytical forms for the first passage time quantities. We proceed
to present our main results, the asymptotic scaling forms of
the first passage time behaviour as function of $H$ and $\Theta$.
These are numerically confirmed in the subsequent Section. We
conclude with a discussion related to reaction kinetics of FBM
particles.

\section{Fractional Brownian motion in wedge domain}

Consider the random walk of a particle confined to a
two-dimensional wedge domain of opening angle $\Theta$, see
Fig.~\ref{wedge}. The particle starts at a position $\mathbf{r}_0$
inside the wedge and moves within the domain until eventually it
hits one of the absorbing boundaries at a time $t$ for the first time.
For this process, we obtain the survival probability
\begin{equation}
\label{surv}
\mathscr{S}_\T(t)=\int_WG(\mathbf{r},t)d\mathbf{r}
\end{equation}
on the wedge domain $W$, and the first passage time density
\begin{equation}
\label{fpt}
\wp_\T(t)=-\frac{d\mathscr{S}(t)}{dt}.
\end{equation}
In Eq.~(\ref{surv}), $G(\mathbf{r},t)$ is the probability density to find
the particle at position $\mathbf{r}$ at time $t$. Due to the presence of
the absorbing boundaries, $G(\mathbf{r},t)$ is not normalised, as expressed
by the survival probability, which decays from $\mathscr{S}_\T(0)=1$ to
$\lim_{t\to\infty}\mathscr{S}_\T(t)=0$.

Two-dimensional FBM is defined as a superposition of independent
FBMs for each Cartesian coordinate \cite{yaglom,JH,vincentref}:
\begin{equation}
\mathbf{r}(t)=\sum_{i=1}^{2}\int_0^t dt'\xi_i^H(t')\hat{x}_i+\mathbf{r}_0,
\end{equation}
where $\hat{x}_i$ is the unit vector in Cartesian direction $i$ ($i=x,y$),
and $\xi^H_i(t)$ is the fractional Gaussian noise. Due to this definition
it is clear that for a quadrant geometry with $\Theta=\pi/2$ the absorption
to either wall decouples. We now first address the case of normal Brownian
motion ($H=1/2$), for which exact results for the survival and first passage
distributions can be found.

\section{First passage process for $H=1/2$}

For normal Brownian motion, the diffusion in a wedge domain is described in
terms of the Green's function $G(r,\theta;t)$, that satisfies the diffusion
equation in polar coordinates,
\begin{equation}
\frac{\partial}{\partial t}G(r,\theta;t)=K_{1/2}\left(\frac{\partial^2}{\partial
r^2}+\frac{1}{r}\frac{\partial}{\partial r}+\frac{1}{r^2}\frac{\partial^2}{
\partial\theta^2}\right)G(r,\theta;t).
\end{equation}
The associated boundary conditions are $G(r,0;t)=G(r,\Theta;t)=0$, representing
the absorbing walls. The solution $G$ is completely specified by the initial
condition $G(r,\theta;0)=\delta(r-r_0)\delta(\theta-\theta_0)/r_0$. At long
times $t$, the solution $G$ can be approximated by \cite{redner}
\begin{equation}
G(r,\theta;t)\simeq\frac{\pi\sin\left(\frac{\pi\theta}{\Theta}\right)}{4K_{1/2}\Theta t}e^{-(r^2+r_0^2)/4K_{1/2}t}I_{\pi/\Theta}
\left(\frac{rr_0}{2K_{1/2}t}\right),
\label{Gsolution}
\end{equation}
where $I_\nu$(z) is the modified Bessel function of the first kind, which can
be expressed by the series expansion
\begin{equation}
I_\nu(z)=(z/2)^\nu\sum_{k=0}^{\infty}\frac{(z^2/4)^k}{k!\Gamma(\nu+k+1)}.
\end{equation}
The survival probability (\ref{surv}) in polar coordinates is then
\begin{equation}
\mathscr{S}_\T(t)=\int_0^\Theta\int_0^\infty rG(r,\theta;t)drd\theta,
\end{equation}
from which the property $\mathscr{S}_\T(0)=1$ follows directly
from the sharp initial condition. Using the approximations
$I_\nu(z)\approx(z/2)^\nu/\Gamma( 1+\nu)$ and
$e^{-r_0^2/4K_{1/2}t}\approx 1$ for the long-time limit in
Eq.~(\ref{Gsolution}), one obtains the scaling expressions
\begin{subequations}
\label{fptdH05}
\begin{eqnarray}
\mathscr{S}_\T(t)&\simeq&\left(\frac{r_0}{\sqrt{K_{1/2}}}\right)^{\pi/\Theta}t^{-\pi/(2
\Theta)},\\[0.2cm]
\wp_\T(t)&\simeq&\frac{\pi}{2\Theta}\left(\frac{r_0}{\sqrt{K_{1/2}}}\right)^{\pi/\Theta}
t^{-1-\pi/(2\Theta)}.
\end{eqnarray}
\end{subequations}
Notably, the wedge angle $\Theta$ enters the scaling exponents
inverse-proportionally; survival and first passage distributions
decay faster for decreasing wedge angle, as it should. We observe
an interesting crossover as function of $\Theta$: as long as
$\Theta \geq\pi/2$ the mean first passage time
\begin{equation}
\mathcal{T}=\int_0^{\infty}t\wp_\T(t)dt
\end{equation}
diverges, $\mathcal{T}\to\infty$. In particular, in the half-space limit
$\Theta=\pi$, the process reduces to the first passage in a one-dimensional,
semi-infinite geometry (along the $y$-axis), and we find the usual
L{\'e}vy-Smirnov scaling $\wp(t)\simeq t^{-3/2}$. When the wedge angle is
smaller, $\Theta<\pi/2$, the mean first passage time is finite, reflecting
the much higher probability to hit one of the two walls of the wedge. Thus,
for $\Theta=\pi/2$, the absorption to either of the two walls can be viewed
as two decoupled Brownian random walks in $x$ and $y$ direction. The resulting
survival probability is then given as the product of two one-dimensional
survival functions corresponding to $\Theta=\pi$:  $\mathscr{S}_{\pi/2}(t)=
\mathscr{S}_{\pi}(t)^2\simeq t^{-1}$, for which the mean first passage time
is logarithmically divergent. Remarkably, the scaling relation
\begin{equation}
\mathscr{S}_{\Theta/2}(t)\simeq\mathscr{S}_{\Theta}^2(t)
\end{equation}
holds more generally for any $\Theta$, see relations (\ref{fptdH05}).

\section{First passage process for $H\neq 1/2$}

Let us now address the case of general Hurst exponent, $H\neq1/2$, for which
no analogue to the above Green's function method is known. While there exist
dynamic equations for FBM in literature \cite{lutz,physica} of the form
\begin{equation}
\frac{\partial}{\partial t}\mathscr{G}(\mathbf{r},t)=K(t)\nabla^2\mathscr{G}(
\mathbf{r},t)
\label{diffusion2}
\end{equation}
with the time-dependent diffusion coefficient $K(t)\sim t^{2H-1}$, this
description cannot fully specify the behaviour of FBM in the presence of
non-natural boundary conditions. To see this, we follow the procedure of
the preceding Section, and find the first passage behaviours
\begin{subequations}
\label{wrong}
\begin{eqnarray}
\mathscr{S}_\T^{\mathscr{G}}(t)&\propto& t^{-\pi H/\Theta},\\
\wp_\T^{\mathscr{G}}(t)&\propto& t^{-1-\pi H/\Theta}.\label{wrongfptd}
\end{eqnarray}
\end{subequations}
While, naturally, the Brownian case $H=1/2$ is consistent with the results of
the previous Section, the results (\ref{wrong}) are inconsistent with Molchan's
result $\wp(t)\simeq t^{H-2}$ for a completely open wedge, $\Theta=\pi$.
Therefore, for arbitrary wedge angle $\Theta$ and Hurst exponent $H$ the
distributions (\ref{wrong}) cannot be correct. In fact, the same inconsistency
is found when one naively applies the method of images to the free space
solution $\left(4\pi K_Ht^{2H}\right)^{-1/2}\exp\left(-x^2/\left[4K_Ht^{2H}
\right]\right)$.

We now argue in favour of a conjecture for the correct scaling forms of the
first passage time quantities. Let us start by recollecting the known special
cases, for which analytical results are available:

(i) When $\Theta=\pi$,
the first passage time quantities must converge to Molchan's result for a
one-dimensional semi-infinite domain \cite{molchan}:
\begin{subequations}
\begin{eqnarray}
\mathscr{S}_{\pi}(t)&\simeq& t^{H-1},\label{sv1d}\\
\wp_{\pi}(t)&\simeq& t^{H-2}, \label{fptd1D}
\end{eqnarray}
\end{subequations}
valid in the long time limit $t\gg1$.

(ii) When $\Theta=\pi/2$, due to the
independence of $x$ and $y$ motion the survival probability, by above argument,
is given as the product of two one-dimensional survival probabilities,
$\mathscr{S}_{\pi/2}=\mathscr{S}_{\pi}^2$, and thus
\begin{subequations}
\label{fptd90}
\begin{eqnarray}
\mathscr{S}_{\pi/2}(t)&\simeq& t^{2H-2},\label{sv90}\\
\wp_{\pi/2}(t)&\simeq& t^{2H-3}.
\end{eqnarray}
\end{subequations}

(iii) For $H=1/2$, we know the analytical results (\ref{fptdH05})
with the full $\Theta$-dependence.

Assuming that the first passage time exponents are simple combinations of the
Hurst exponent $H$ and the wedge angle $\Theta$, to satisfy the above three
special cases we conjecture the unique scaling form
\begin{subequations}
\begin{eqnarray}
\mathscr{S}_\T(t)&\simeq& t^{\pi(2H-2)/(2\Theta)},\label{S}\\
\wp_\T(t)&\simeq& t^{-1+\pi(2H-2)/(2\Theta)}.
\end{eqnarray}\label{fptd}
\end{subequations}
These results imply that the survival decays faster for slower diffusion
(i.e., lower value of $H$), in analogy with Molchan's result and findings
for the barrier crossing of FBM \cite{oleksii}. Note that the inconsistent
results (\ref{wrong}) based on the dynamic equation (\ref{diffusion2}) follow
from our results through the substitution $H\rightarrow1-H$.

\section{Simulations}

To numerically verify the conjecture (\ref{fptd}) for the survival
probability and the first passage time density, we performed
simulations for various wedge angles and Hurst exponents, and
compared the results with Eqs.~(\ref{fptd}). For this comparison,
we here focus on the survival probability, which generally shows
better statistics.

Our simulations procedure is as follows. A wedge of opening angle $\Theta$ is
constructed by imposing two absorbing walls at $y=0$ and $y=(\tan\Theta)x$ in
the $x$-$y$ plane (Fig.~\ref{wedge}(a)). At $t=0$, the particle is located at
$\mathbf{r}_0=(r_0\cos\frac{\Theta}{2},r_0\sin\frac{\Theta}{2})$ on the line
bisecting the wedge. For $t>0$, the particle undergoes two-dimensional FBM in
discrete time indexed by $n$, following the rule
\begin{eqnarray}
\nonumber
&&\hspace*{-0.8cm}\mathbf{r}(t_n)=\\
&&\hspace*{-0.8cm}\left(\sum_{m=1}^{n}\xi_1^H(t_{m})+r_0\cos\frac{\Theta}{2},
\sum_{m=1}^{n}\xi_2^H(t_{m})+r_0\sin\frac{\Theta}{2}\right).
\end{eqnarray}
Here, the two independent fractional Gaussian noise functions $\xi^H_1(t)$
and $\xi^H_2(t)$ are generated by the Hosking method \cite{hosking,JH}. The
simulation is terminated when the particle escapes the wedge domain for the
first time. This is the survival time for the particle in the given run.
From repeated runs we construct the survival probability $\mathscr{S}_\T(t)$ as
the accumulated number of particles surviving until time $t$. For each given
wedge angle and Hurst exponent, the survival probability $\mathscr{S}_\T(t)$ was
obtained from 100,000 runs and for initial radius $r_0=0.25$.

\begin{figure}[tb]
\centering
\onefigure[width=8.8cm]{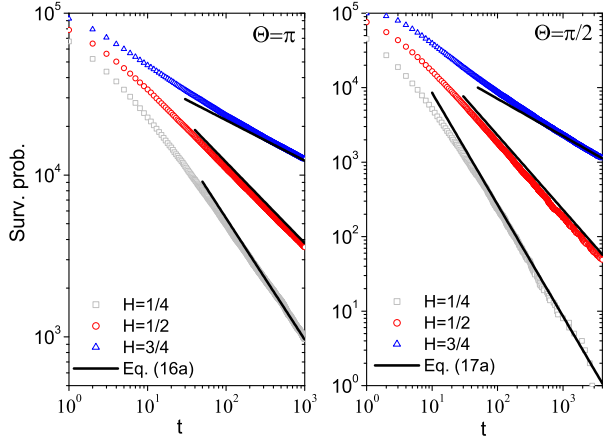}
\caption{Survival probability $\mathscr{S}_\T(t)$ for wedge angles $\Theta=\pi$
(Left) and $\Theta=\pi/2$ (Right) with the Hurst exponents $H=1/4$, 1/2, and
3/4 (from bottom to top). The lines represent the expected scaling behaviours,
Eqs.~(\ref{sv1d}) and (\ref{sv90}).}
\label{Stpi}
\end{figure}

To confirm that our simulations procedure produces correct results
for the survival probability, we first consider the two special
cases of wedge angles $\Theta=\pi$ and $\pi/2$, and compare them
to the predicted scaling behaviours (\ref{sv1d}) and (\ref{sv90}).
Fig.~\ref{Stpi} shows on the left the distributions of
$\mathscr{S}_\T(t)$ obtained from simulations of a fully open
wedge, $\Theta=\pi$, for Hurst exponents $H=1/4$ (subdiffusion),
1/2 (normal diffusion), and 3/4 (superdiffusion). Clearly, the
distributions follow the predicted scaling behaviours for this
case corresponding to the one-dimensional, semi-infinite domain.
Similarly, for the case of a rectangular wedge ($\Theta=\pi/2$) on
the right of Fig.~\ref{Stpi} we find excellent verification of our
simulations method, compared to the predicted behaviour
$\mathscr{S}_{\pi/2}(t)\sim t^{2H-2}$ for all values of $H$.

\begin{figure*}
\centering
\includegraphics[width=18cm]{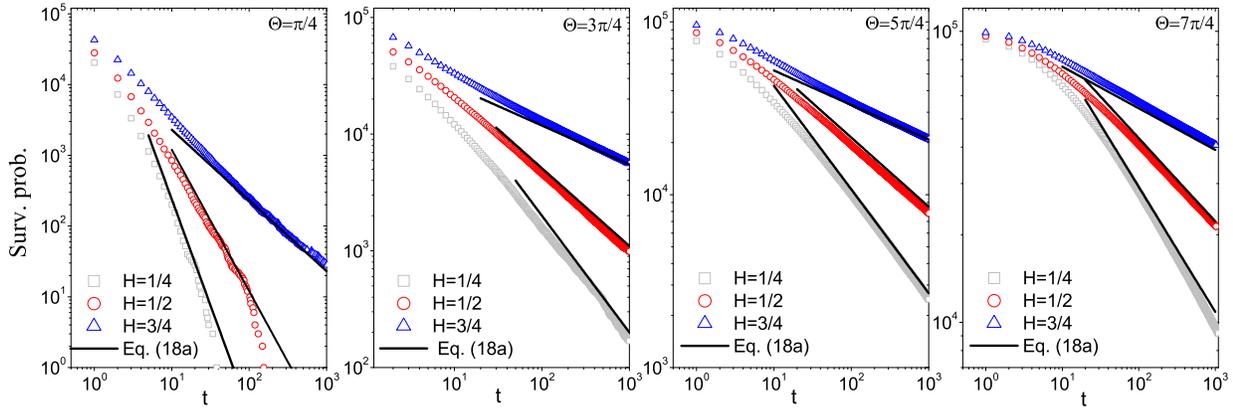}
\caption{Survival probability $\mathscr{S}_\T(t)$ for wedge angles $\Theta=\pi/4$,
$3\pi/4$, $5\pi/4$, and $7\pi/4$, with Hurst exponents $H=1/4$, 1/2, and 3/4
(from bottom to top). The corresponding expected scaling behaviours
(cf.~Eq.~\eqref{S}) are depicted by the full lines. }
\label{Stpi2}
\end{figure*}

We now investigate whether our conjecture (\ref{S}) for the
survival probability also holds for other wedge angles. In our
simulation, the survival probabilities for wedge angles
$\Theta=\pi/4$, $3\pi/4$, $5\pi/4$, and $7\pi/4$ were
investigated, for Hurst exponents $H=1/4$, 1/2, and 3/4. As
demonstrated in Fig.~\ref{Stpi2}, the simulated survival
probabilities remarkably well follow our conjecture (\ref{S}) for
all wedge angles and Hurst exponents. Particularly, the scaling
exponents of the survival indeed decreases for growing values of
the Hurst exponent $H$, in line with Molchan's result and findings
for the barrier crossing behaviour of FBM.

\section{Discussion}

We studied the first passage properties of two-dimensional FBM
confined to a wedge domain with absorbing boundaries. Starting
from special cases, for which exact long-time scaling expressions
can be analytically derived, we obtained functional dependencies
of first passage time exponents on the Hurst exponent $H$ and
wedge angle $\Theta$. By numerical analysis we confirmed this
conjecture. We believe that our results are an important
additional brick in the construction of a clear picture of the
elusive FBM process. The obtained forms (\ref{fptd}) of the first
passage time quantities imply that the mean first passage time
of a particle from a wedge domain across the domain walls,
$\mathscr{T}$, diverges for wedge angles
$\Theta\geq\Theta_c$. The critical angle corresponds to
$\Theta_c=\pi(1-H)$.

From the above results one may obtain some insight into the
reaction kinetics of three particles in one-dimensional space,
driven by FBM (Fig. \ref{wedge}(b)): \emph{(1) Surrounded prey.}
Let us consider the diffusion-limited reaction of the central
particle with either of the two comrade particles. The problem can
be transformed such that one particle is confined between two
moving absorbing boundaries \cite{redner}. If $x_i(t)$ denotes the
position of the $i$th particle at time $t$, the survival condition
of the central particle is $x_1(t)<x_2(t)$ and $x_2(t)<x_3(t)$ for
particles sharing the same diffusion properties ($H$ and $K_H$).
We regard $x_i(t)$ as the $i$th coordinate of a single
particle, that diffuses in three-dimensional space. Then, the
survival criterion is understood such that this particle diffuses
inside a domain limited by absorbing planes at $x_1=x_2$ and
$x_3=x_2$. This is equivalent to the first passage problem for
two-dimensional motion in a wedge domain, the wedge angle given by
the two intersecting planes, i.e., $\Theta_{\mathrm{mid}}=\pi/3$
(see Ref. \cite{redner}). Thus, from the distribution (\ref{S}) we
find that the survival probability for the central particle has
the asymptotic form $\mathscr{S}_{\mathrm{mid}}(t)\sim
t^{3(H-1)}$. Contrary to our naive expectation, the above scaling
exponent shows that superdiffusive particles (with $H>1/2$)
survive longer than ordinary Brownian particles ($H=1/2$), and
subdiffusive particles have a better chance to meet each other.
These characteristics will lead to significant differences in the
total reaction amount, compare Figs. \ref{Stpi} and \ref{Stpi2}.
Intriguingly the mean reaction time $\mathscr{T}$ is finite for
$H<H^*=2/3$, meaning that reactions always occur for subdiffusive
motion, similar to the considerations in Ref.~\cite{guigas}.
\emph{(2) Chased prey.} For the case that one of the corner
particles ($x_3$, say) is chased by the two other diffusing
particles ($x_1$ and $x_2$), the problem is reformulated to the
task of finding the survival probability for the wedge domain of
angle $\Theta_{\mathrm{end}}=2\pi/3$. Hence for identical
particles the probability that one corner particle survives until
time $t$ scales as $\mathscr{S}_{\mathrm{end}}(t)\sim
t^{3(H-1)/2}$. In this case the reaction is slower than that of
the surrounded particle. Note the scaling relation
$\mathscr{S}_{\mathrm{mid}}(t)\simeq\mathscr{S}^
2_{\mathrm{end}}(t)$, such that the reaction of a particle
surrounded by two others corresponds to the product of two
independent corner particle reactions.

We note that closer inspection of the simulated survival probabilities
shows that in the non-Markovian case ($H\neq 1/2$) the first passage
process exhibits a somewhat intricate scaling behaviour: the long-time
behaviours (\ref{fptd}) are preceded by another distinct scaling law
at intermediate times. This behaviour is particularly noticeable for the
cases of wedge angle $\Theta\leq\pi$, for which first passage events occur
easily at short and intermediate times. The apparent intermediate scaling
behaves differently, depending on the diffusion characteristics. For
subdiffusion, the intermediate scaling exponent is smaller than the
long-time exponent $\frac{\pi}{\Theta}(1-H)$ and larger for superdiffusion.

The first passage behaviour displayed by FBM is different from results derived
from the diffusion equation with time-dependent diffusion coefficient. Latter
would lead to an inconsistent $H$-dependence, corresponding to the replacement
$H\to1-H$ in the correct forms (\ref{fptd}). Similar inconsistency
occurs if the method of images were applied to construct the solution in the
presence of non-natural boundary conditions. We note in
passing that the substitution $H\to1-H$ in our results for $\Theta=\pi$ leads
to the scaling for the first passage behaviour found in the analysis of a
generalised Langevin equation in Ref.~\cite{taloni}.

The somewhat counterintuitive behaviour that smaller $H$ implies
faster decay of the first passage time density, may be connected to
the fact that FBM is fuelled by external noise, that is not
balanced by friction. FBM, that is, does not obey the
fluctuation-dissipation relation, in contrast to the generalised
(fractional) Langevin equation \cite{lutz,physica,goychuk} that describes FBM in
conditions
close to thermal equilibrium, when the fluctuation-dissipation
theorem is valid. We will investigate the latter behaviour in a
separate work.

We finally note that while for a compact process in one dimension
the first passage is identical to the first arrival to a given
position, these processes are no longer necessarily equal to each
other in higher dimensions. This may also be the reason why the
Wilemskii-Fixman approximation reproduces Molchan's result
$\wp(t)\simeq t^{H-2}$ in one dimension, but delivers a different
result in higher dimensions.

\acknowledgments

We thank Michael Lomholt and Igor Sokolov for helpful discussion. Financial
support from the Academy of Finland (FiDiPro scheme) and the European
Commission through MC IIF Grant No. 219966 LeFrac is gratefully
acknowledged.

\section{Appendix}

The survival probabilities presented in Fig.~\ref{Stpi2} were obtained with
the same initial starting point and number of simulation runs for consistency.
Depending on the wedge angle and Hurst exponent, however, some cases are
more problematic to show reliable long-time scaling properties within the time
window we used. Here we present supplementary results for these cases.

\begin{figure}
\centering \onefigure[width=8.8cm]{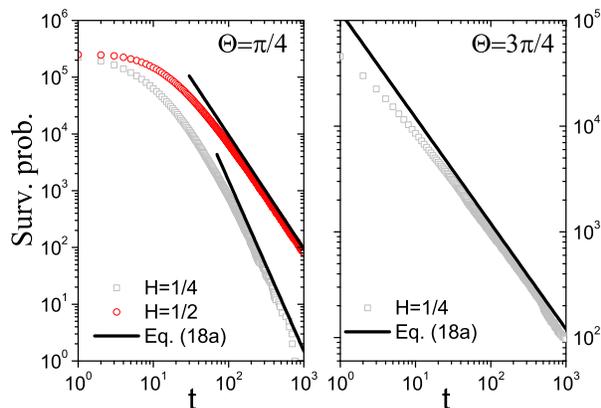} \caption{(Left)
Survival probability $\mathscr{S}_\T(t)$ for wedge angle
$\Theta=\pi/4$ and Hurst exponents $H=1/4$ (below) and 1/2 (above).
Here $r_0=2.0$, and the number of simulation runs is
250,000. (Right) Survival probability $\mathscr{S}_\T(t)$ at wedge
angles $\Theta=3\pi/4$ for the Hurst exponent $H=1/4$. Here
$r_0=0.10$, and number of simulation runs is $100,000$. The solid
lines are the expected power-laws predicted by Eq.~\eqref{S}.}
\label{supp1}
\end{figure}

(i) Fig.~\ref{supp1} depicts the case for $\Theta=\pi/4$ with
$r_0=2.0$ (i.e., eight times the value used in the main text) and
250,000 runs. Now the survival probabilities for $H=1/4$ and $1/2$
exhibit satisfactory long-time scaling.

(ii) To obtain extended long-time scaling behaviour for $H=1/4$
and wedge angle $\Theta=3\pi/4$, the initial distance from
the origin was decreased to $r_0=0.10$. The obtained result in
Fig.~\ref{supp1} shows good agreement with Eq.~\eqref{S}.

\end{document}